\documentclass[10pt,conference]{IEEEtran}
\IEEEoverridecommandlockouts
\usepackage{comment}
\usepackage{cite}
\usepackage{amsmath,amssymb,amsfonts}
\usepackage{algorithmic}
\usepackage{graphicx}
\usepackage{textcomp}
\usepackage{xcolor}
\usepackage{tabularx}
\usepackage{siunitx}
\usepackage{balance}
\usepackage[T1]{fontenc}

\newcolumntype{B}{X}
\newcolumntype{R}{>{\hsize=.45\hsize}X}
\newcolumntype{Y}{>{\hsize=.25\hsize}X}

\begin{document}

\title{Documentation Practices in Agile Software Development: A Systematic Literature Review}

\author{
    \IEEEauthorblockN{Md Athikul Islam}
    \IEEEauthorblockA{
        \textit{Department of Computer Science} \\
        \textit{Boise State University} \\
        Boise, ID, USA \\
        mdathikulislam@u.boisestate.edu
    }
    \and
    \IEEEauthorblockN{Rizbanul Hasan}
    \IEEEauthorblockA{
        \textit{Department of Computer Science} \\
        \textit{Boise State University} \\
        Boise, ID, USA \\
        rizbanulhasan@u.boisestate.edu
    }
    \and
    \IEEEauthorblockN{Nasir U. Eisty}
    \IEEEauthorblockA{
        \textit{Department of Computer Science} \\
        \textit{Boise State University} \\
        Boise, ID, USA \\
        nasireisty@boisestate.edu
    }
}

\maketitle

\begin{abstract}
\textit{Context}: 
Agile development methodologies in the software industry have increased significantly over the past decade. Although one of the main aspects of agile software development (ASD) is less documentation, there have always been conflicting opinions about what to document in ASD.
\textit{Objective}: 
This study aims to systematically identify what to document in ASD, which documentation tools and methods are in use, and how those tools can overcome documentation challenges.
\textit{Method}: 
We performed a systematic literature review of the studies published between 2010 and June 2021 that discusses agile documentation. Then, we systematically selected a pool of 74 studies using particular inclusion and exclusion criteria. After that, we conducted a quantitative and qualitative analysis using the data extracted from these studies.
\textit{Results}: 
We found nine primary vital factors to add to agile documentation from our pool of studies. Our analysis shows that agile practitioners have primarily developed their documentation tools and methods focusing on these factors. The results suggest that the tools and techniques in agile documentation are not in sync, and they separately solve different challenges.
\textit{Conclusions}: 
Based on our results and discussion, researchers and practitioners will better understand how current agile documentation tools and practices perform. In addition, investigation of the synchronization of these tools will be helpful in future research and development.

\end{abstract}

\begin{IEEEkeywords}
Software Engineering; Agile Software Development; Documentation; Systematic Literature Review
\end{IEEEkeywords}

\section{Introduction}
\label{sec:Introduction}
Software documentation is an integral part of software development. It works as a communication medium between developers of a team and is utilized as an information repository by maintenance engineers \cite{sommerville2001software}. Documentation elucidates how the system is structured, its functionalities, and the design rationale \cite{5287001}. Software documentation should be included as part of software development and is sometimes called “common sense” \cite{10.1145/585058.585065}. Even outdated document serves a purpose and may be helpful \cite{10.1145/585058.585065}. However, the incomplete, wrong, clumsy, abstruse, outdated, and inadequate document often leads to the unpopularity of software documentation among the developers \cite{8811931}. Regardless of the application type, almost all medium to large software projects produces a certain amount of documentation \cite{sommerville2001software}. 

ASD is an iterative approach that helps teams deliver value to their customers faster and more efficiently. It is immensely applied in both industry and academia \cite{iet:/content/journals/10.1049/iet-sen_20070038}. This incremental approach maintains a strong focus on project goals and customer involvement. Over the past decade, numerous software developers have adopted the ASD model \cite{heck2017framework}. This concept of agile is gained by different agile methodologies like Scrum, eXtreme Programming (XP), Crystal, Lean, Dynamic Systems Development Method (DSDM), and Feature-Driven Development  (FDD). Documentation has a lower priority than working software in Agile practices \cite{fowler2001agile}. Some agile practitioners consider code as its documentation. As a result, the information that is recorded in documentation (documented information, henceforth) may not be well maintained, resulting in inadequate information for the team to understand development tasks \cite{8049141}. Another reason for less focus on documentation is that it consumes time that could have been allocated in development~\cite{5287001}.  

As modern software systems are complicated, developers revisit the system  more often. Trivial maintenance work is assigned to junior developers who have little experience with the code. As a result, lack of documentation hinders inter-team building and knowledge loss \cite{7352703}. In addition, users of a software product expect quality results \cite{8584455}. Considering these scenarios, agile documentation plays a significant role in ASD \cite{5287001, 8595625}. Moreover, offshore agile development is widespread nowadays, and documentation is one of the key factors to make offshore agile development successful \cite{4777400}. Therefore, developers should document the features of each iteration to help future developers refactor them into smaller tasks.  

The practitioners working on agile documentation have implemented different documentation strategies to overcome the issues raised by the lack of documentation. They have identified different key elements that developers should document such as user stories, functional requirements, source code, etc \cite{6824127, 10.1145/3383219.3383245, KHAMIS201319}. They have also developed various tools and models such as wikis, simul loco, doxygen, etc to document these elements effectively \cite{ambler2002agile, 6263161, 4151761}. 
In this paper, we focus on understanding the pivotal information to document in agile and the existing techniques and tools that result in document optimization. 
We conducted a systematic literature review from existing research and analyzed them to answer our research questions mentioned in section~\ref{sec:Methodology}.
Our findings will benefit any agile practitioners and developers to optimize their documentation effort and help researchers in further study. 




\section{Research Methodology}
\label{sec:Methodology}
In our research methodology, we followed the directions proposed by Kitchenham and Charters \cite{kitchenham2007guidelines}. We have divided our research review into four steps, namely planning, conducting, and reporting the review results.

\subsection{Planning}

We have planned this review by confirming the need for such a review and have proposed our research questions accordingly. Our planning phase includes search strategy, search string, and inclusion/exclusion criteria.

\subsubsection{Research questions}

We pose the following research questions to drive this study.

\begin{itemize}
    \item \textbf{RQ1: Which information do agile practitioners document? }
    \item \textbf{RQ2: Which documentation generation tools and methods do agile practitioners use?}
    \item \textbf{RQ3: How can the tools and methods overcome the documentation challenges in agile software development?}	
\end{itemize}

\subsubsection{Search strategy}
After defining the need for this systematic review and research questions, we started to carry out the formulation of a search strategy based on the guidelines provided by Kitchenham and Charters \cite{kitchenham2007guidelines}. In Table \ref{table:searchstring} we broke down the question into individual facets i.e. population, intervention, and constructed search string using Boolean ANDs and ORs. We initially fetched studies from the electronic databases and then explored them through reference searches (snowballing) to seek other meaningful studies. After that, we applied our inclusion and exclusion criteria to the fetched studies involving a different number of researchers, as explained in Section \ref{sssec:inc_exc}.

\subsubsection{Search criteria}
The search criteria used for this review consist of three parts - C1, C2, and C3, defined as follows:
\begin{itemize}
    \item{C1: We constructed the C1 string, which enables the keyword agile either in the title or abstract.}
    \item{C2: The C2 string is made up of keywords such as document or document* either in title or abstract.}
    \item{C3: We constructed the C3 part, which enables keywords such as tools or document* tools either in the title or abstract.}
\end{itemize}

The boolean expression: C1 \textit{AND} C2 \textit{OR} C3

We provided our search string in Table \ref{table:searchstring}.

Another key thing to note is that we have filtered out the result fetched from the search query by applying the checkbox feature of the IEEE Xplore database. In this case, we filtered out publication topics such as internet, organizational aspects, computer-aided instruction, DP industry, computer science education, educational courses, knowledge management, mobile computing, security of data, business data processing, and teaching not relevant to our research.

\begin{table}[htbp]
\caption{Search string}
\label{table:searchstring}
\begin{center}
\def\arraystretch{1.5}%
\begin{tabular}{ |c| } 
 \hline
 Search string \\ 
 \hline
((("Document Title": "agile" OR "Abstract": "agile") \\
AND ("Document Title": "document*" OR "Abstract": \\
"document*")) OR ("Document Title": "document* \\
tools" OR "Abstract": "document* tools"))\\
 \hline
\end{tabular}
\end{center}
\end{table}

\subsubsection{Inclusion and exclusion criteria}
\label{sssec:inc_exc}

As per the guidelines of Kitchenham and Charters \cite{kitchenham2007guidelines} we have set inclusion and exclusion criteria based on our research questions. 
Here, we only considered papers that are in English, published in conferences and journals, and published within the time frame 2010 - 2021. The published papers should describe the agile documentation approach, tools, or knowledge relevant to our RQs. Therefore, we did not include any opinion, viewpoint, keynote, discussions, editorials, comments, tutorials, prefaces, anecdote papers, and presentations. In addition, we excluded the papers that did not discuss agile documentation or agile documentation tools but may discuss agile software development methods as a side topic. 

\subsection{Conducting the review}

Once we agreed on the protocol, we started our review properly. This section discusses the findings of our search and extracted data from relevant databases and sources.

\subsubsection{Study search and selection}

We searched the IEEE Xplore database against our search query and search criteria and fetched 206 studies. In round 1, the first author immensely analyzed the titles and abstracts of the fetched studies based on the inclusion criteria. After the first round, we came out with 77 papers, and most of these studies covered all the inclusion criteria. A critical part was to ensure the papers did not come from opinions, discussions, editorials, comments, tutorials, prefaces, and presentations as per the exclusion criteria. In round 2, we inspected the full-text review of the papers based on all inclusion and exclusion criteria. 
We read the papers fully a few times where there were disagreements and required consensus. We excluded 23 papers based on the exclusion criteria.
Finally, to satisfy the inclusion of relevant primary studies as much as possible, we performed backward and forward snowballing following the guideline provided by Wohlin \cite{wohlin2014guidelines} and included 20 more papers in our list of primary studies. The final pool of selected papers was 77. 


\subsubsection{Data extraction}

We followed the data extraction strategy of  Kitchenham and Charters \cite{kitchenham2007guidelines} and came up with a data extraction form that we designed to collect all the information needed to address the review questions. We set a few quality evaluation criteria, such as

\begin{itemize}
    \item{How well was data collection carried out?}
    \item{Is the research design defensible?}
    \item{How much are the findings credible? }
    \item{Is the research scope well addressed?}
\end{itemize}

In addition to the RQs and quality evaluation criteria, the form included the data as (i) name of the reviewer, (ii) date of data extraction, (iii) title, (iv) authors, (v) journal, (vi) publication details, (vii) future work, (viii) limitations, (ix) year of publication, (x) methodology, (xi) data analysis, (xii) validation technique, (xiii) relevancy and xiv) space for additional notes. This data extraction was performed independently by the first and second authors.

\subsubsection{Data synthesis}

After the data extraction, we combined and summarized the results of the included primary studies according to the guidelines of Kitchenham and Charters \cite{kitchenham2007guidelines}. Our data synthesis includes both quantitative and descriptive. For the descriptive synthesis, we tabulated the data based on research questions. In this case, we synthesized what type of information and tools are used in agile documentation. On the other hand for quantitative synthesis, we again developed a tabular form for research questions. These tables were structured to highlight similarities and differences between study outcomes. Later the data of these tables were represented by bar charts and pie charts.

\section{Results}
\label{sec:Results}
In this section, we present our findings. We address each research question from RQ1 to RQ3. 

\subsection{\textbf{(RQ1) Which information do agile practitioners document?}} 

\begin{table*}[htbp]
\caption{Key elements to add in agile software documentation}
\label{table:keyfactors}
\begin{center}
\def\arraystretch{1.3}
\begin{tabularx}{\textwidth}{S R B}
\hline
\textbf{No.} & \textbf{Information to document}  & \textbf{Studies that reported the information} \\
\hline
1 & User stories & \cite{6824127}, \cite{10.1145/3383219.3383245}, \cite{6113071}, \cite{7765512}, \cite{6189437}, \cite{8054887}, \cite{8104703}, \cite{8560921}, \cite{9609208},  \cite{6149538}, \cite{7352703}, \cite{5700043}   \\
\hline
2 & Functional requirements & \cite{6824127}, \cite{6614746}, \cite{9609208}, \cite{6149538}, \cite{8595625} \\
\hline
3 & Non-functional requirements & \cite{6824127},  \cite{9226305}, \cite{10.1145/3383219.3383245}, \cite{7059146}, \cite{8104703},  \cite{7320450}, \cite{10.1145/2961111.2962616} \\
\hline
4 & Source code & \cite{6824127}, \cite{KHAMIS201319}, \cite{10.1007/978-3-642-13881-2_7}, \cite{4151761}, \cite{6263161},   \cite{10.1007/978-3-642-21043-3_49}  \\
\hline
5 & UI structure &  \cite{10.1145/3383219.3383245}, \cite{8054887}, \cite{8104703}, \cite{9609208}, \cite{7352703} \\
 \hline
6 & Technical debt & \cite{9226305} \\
 \hline
7 & System architecture & \cite{6824127}, \cite{7765512}, \cite{6614746}, \cite{7352703}  \\
 \hline
8 & API reference guides & \cite{7961538}, \cite{9609208},  \cite{8103450} \\
 \hline
9 & Test specification & \cite{6824127},  \cite{isoieee}, \cite{8595625} \\
 \hline
\end{tabularx}
\end{center}
\end{table*} 

Table \ref{table:keyfactors} summarizes our findings of documented information. Our primary pool of studies consisted of interviews, surveys, case studies, experiments, and statistical analysis. Many agile practitioners, graduate students, and software engineers directly participated in these studies. We collected the findings from these studies and grouped them. The following sections briefly describe each element of Table \ref{table:keyfactors}.

\subsubsection{User stories} 

User stories function as the shortcut for more formal documentation and require more details \cite{8590164}. On the other hand, they represent the small, concise user-driven features and hence need to be documented \cite{8560921}.

\subsubsection{Functional requirements} 

End-users define these requirements and expect them as facilities when they interact with the system. Therefore, these functionalities focus more on the technical aspects that need to be implemented \cite{9609208}.

\subsubsection{Non-functional requirements} 

The success of an agile project depends on the non-functional requirements, which are also referred to as quality requirements \cite{9226305}. These requirements are quality constraints that must be satisfied by the system. So, failure to meet these requirements compromises the entire system and makes it useless \cite{7059146}.

\subsubsection{Source code} 

The source code should be documented for traceability and deriving the code that does not perform~\cite{6824127}.

\subsubsection{UI structure} 

These are wireframes and sometimes are documented externally. However, they can be the basic layouts of application screens and the product owner usually provides these wireframes \cite{10.1145/3383219.3383245}.

\subsubsection{Technical debt} 

If a system needs fixes or updates, it is best to document them currently. Similarly, it is best to document them in the case of technical debt. As a result, future developers will be aware of that technical debt. Therefore, all instances of technical debt should be documented \cite{HOLVITIE2018141}. 

\subsubsection{System architecture}

Some documentation tools have evolved to document architecture \cite{6824127, 6614746}. Architecture works as the backbone of the entire system \cite{6824127}. Documenting architecture is one of the most complicated and challenging parts of software development \cite{6614746}. 

\subsubsection{API reference guides}

To understand the usage and integration of API, APIs should have comprehensive documentation. Good API reference guides make the APIs easier to maintain and helps onboard new developers to the team~\cite{8103450}. 

\subsubsection{Test specification}

It is tough to demonstrate large-scale systems solely using test cases in the industry. Testing needs more mature documentation to keep track of the test cases, user scenarios, and bugs \cite{6824127}.

\subsection{\textbf{(RQ2) Which documentation generation tools and methods do agile practitioners use?}}

\begin{figure}[htbp]
\centerline{\includegraphics[width=8.5cm]{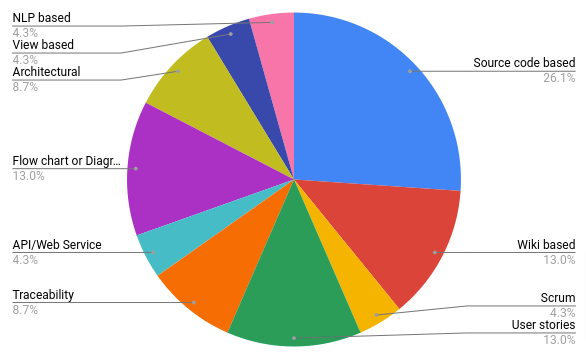}}
\caption{Categorisation of documentation tools/methods}
\label{figure:categorizationoftools}
\end{figure}

In order to answer this research question, we first explored the current tools and methods used in ASD from our primary pool of studies and found a total of 23 tools and methods. Next, we categorized them based on their document type and found ten categories. Table \ref{table:toolsandmethods} lists categories alongside their tools and Figure \ref{figure:categorizationoftools} shows the percentage of tools under each category. We also listed the studies that reported the tools and methods. Moreover, we mentioned the role of each tool or method in the right-most column of the table.

\begin{table*}[htbp]
\caption{Tools and methods used in agile software documentation}
\label{table:toolsandmethods}
\begin{center}
\def\arraystretch{1.3}
\begin{tabularx}{\textwidth}{S Y Y Y B}
\hline
\textbf{No.} & \textbf{Documentation type} & \textbf{Tools/Methods} & \textbf{Studies that reported the practice} & \textbf{Role} \\
\hline
1 & Source code based documentation & simul loco & \cite{6263161} & Creates a graphical comment layer over source code that can contain any resource for documentation \cite{6263161}. \\
 & & doxygen & \cite{4151761}, \cite{10.1007/978-3-642-13881-2_7} & JavaDoc or docstring like documentation tool for Java, C++, Python, and other languages \cite{4151761}.\\
 & & JavadocMiner & \cite{10.1007/978-3-642-13881-2_7}, \cite{10.1007/978-3-642-21043-3_49}, \cite{KHAMIS201319} & Wrapper of JavaDoc and provides quality assessments and recommendations on how Javadoc comments can be improved \cite{10.1007/978-3-642-13881-2_7}.\\
 & & GitHub plus markdown & \cite{8804485} &  Agile 'Docs Like Code' solution that puts documentation close to the code while software development tools and techniques are applied \cite{8804485}.\\
 & & Graphical UML class models & \cite{7967949}, \cite{6298089}, \cite{6832301}, \cite{7059146}, \cite{8449554}, \cite{5513837}, \cite{6895218}, \cite{8782418} & Graphical UML class models for source code in continuous agile development \cite{7967949}.\\ 
 & & XML\_DocTracker & \cite{7885872} & Produces the software requirements specification (SRS) from the source codes if the SRS did not exist
for that particular software \cite{7885872}.\\
 \hline
 2 & Wiki based & Wikis & \cite{ambler2002agile}, \cite{silveira2005wiki}, \cite{5978877} & Provides a collaborative environment that people use to co-author HTML-based information \cite{ambler2002agile}. \\
 & & sprintDoc & \cite{7521550} & Works with issue tracker, wikis, VCS, IDE \cite{7521550}.\\
 & & XSDoc & \cite{7521550}, \cite{10.1145/1104973.1104980}, \cite{aguiar2003xsdoc} & Minimizes the development–documentation gap by making documentation more user-friendly and attractive to developers \cite{10.1145/1104973.1104980}.\\ 
\hline
3 & Scrum & Scrumconix & \cite{7577440} & Composition of scrum and ICONIX methods that uses a lightweight approach to document in AGSD environments \cite{7577440}. \\
\hline
4 & User stories & COSMIC method & \cite{6113071}, \cite{7000106} & Assesses the quality of the documentation by analyzing how functional processes are documented in the requirements, such as in the user stories \cite{6113071}. \\
 & & CMMI & \cite{6968220}, \cite{6209192} & Divides the goals into two parts and emphasizes that the product must be delivered on time and meet all of the specified standards \cite{6968220}.\\ 
 & & LAQF & \cite{8966716} & Lightweight document oriented
reusable agile quality framework \cite{8966716}.\\ 
\hline
5 & Traceability & TraceMan & \cite{6824127} & Traces among user stories, traditional requirements documents, test specifications, architecture design, and source code \cite{6824127}.\\
 & & Trace++ & \cite{7765512} & Inherits traditional traceability relationships in order to support the transition from traditional to ASD \cite{7765512}.\\ 
\hline
6 & API/Web Service & Docio & \cite{7961538} & Documents API Input/Output with parameter and response type \cite{7961538}. \\
\hline
7 & Flow chart or Diagrams & Flowgen & \cite{6975637} & Generates flowcharts from annotated C++ source code and high-level UML activity diagrams, one for each function or method in
the C++ sources \cite{6975637}. \\
 & & CLARET & \cite{8491148} & Facilitates functionality to create the use case specifications using natural language \cite{8491148}.\\
 & & Ticket-commit network chart (TCC)
 & \cite{8048905, saito2018discovering} & Visually represents time-series commit activities with issued tickets \cite{saito2018discovering}.\\

\hline
8 & Architectural & Abstract specification tool & \cite{6614746} & Contains the most relevant and essential information on the architecture solution \cite{6614746}.  \\
 & & Active Documentation Software Design (ADSD)
 & \cite{Rubin2011} & Provides an architectural design in which domain knowledge is represented explicitly and is isolated from other segments of code \cite{Rubin2011}.\\
\hline
9 & View based & View-based software documentation & \cite{1607376} & Utilizes existing software modeling techniques and improves the current methods of software documentation \cite{1607376}.  \\
\hline
10 & NLP based & JavadocMiner & \cite{10.1007/978-3-642-21043-3_49} & Provides a complete environment for embedding NLP into software development \cite{10.1007/978-3-642-21043-3_49}.  \\
\hline
\end{tabularx}
\end{center}
\end{table*}

\subsubsection{Source code based documentation} 
Source code-based documentation can mitigate certain risks \cite{8991647}. This category consists of 6 tools, and these tools mainly focus on how software practitioners can generate documentation from source code comments. These tools cover the functionalities of some of the popular documentation generation tools like JavaDoc or Docstring and offer some additional features \cite{10.1007/978-3-642-13881-2_7}.

\subsubsection{Wiki based documentation} 
When developers consider straightforward and flexible documentation options in agile, they consider wiki-based documentation in the first place because the primary goal of the wiki is to minimize the development–documentation gap by making documentation more convenient and attractive to developers \cite{7521550}. For example, sprintDoc and XSDoc are tools based on wikis and can be integrated with other tools such as the VCS IDE.

\subsubsection{Scrum} 
Scrum is a lightweight framework for agile development and is very popular. Scrumconix supporting scrum proved to be a valuable and lightweight tool to document and understand a software project \cite{7577440}.

\subsubsection{User story} 
The user stories explain how the software will work for the users and provide an essential source for the design of the software according to user needs \cite{isoieee}. Methods such as the COSMIC method \cite{6113071, 6149538} can measure the quality of user stories to generate high-quality documentation.

\subsubsection{Traceability} 
Traceability tools like Trace++ support the transition from traditional to agile methodologies. They offer traceability between documents generated during conventional software development and agile methods \cite{7765512}. In addition, TraceMan provides traces to critical agile artifacts \cite{6824127}.

\subsubsection{API/Web service} 
In this category, we found a tool called Docio which can generate API documents with I/O examples \cite{7961538}. This tool is more like the popular REST API documentation generation tool Swagger but only supports the C programming language \cite{surwase2016rest}.

\subsubsection{Flow chart} 
A few tools emphasize providing meaningful graphical diagrams either based on source code or requirements \cite{6975637, 8048905}. Flowgen, CLARET,  and TCC are some of the tools in this category.

\subsubsection{Architectural} 
Experts ascertained the lack of documentation and architectural design in agile projects \cite{6225956}. Abstract specification tool in this category assists the architects in organizing relevant information regarding the architecture while creating design and architecture blueprints, thus reducing the effort of documentation \cite{6614746}. Also, Active Documentation Software Design (ADSD) is an approach that enables incorporating domain documentation to agile development, while having the processes adaptive \cite{Rubin2011}.

\subsubsection{View based} 
View-based software documentation enables different perspectives on the software system and enables the explicit and simultaneous modeling of all of those viewpoints as views in the documentation \cite{1607376}.

\subsubsection{NLP-based} 
Researchers are aware of integrating modern NLP-based techniques and tools into the source code comments where documentation is only available in the form of source code comments. As a result, these tools directly contribute to determining the quality of documentation. JavadocMiner is one of such NLP-based tools that developers can easily embed with Eclipse IDE \cite{10.1007/978-3-642-21043-3_49}.

\subsection{\textbf{(RQ3) How can the tools and methods overcome the documentation challenges in agile software development?}}

\begin{figure}[htbp]
\centerline{\includegraphics[width=9cm]{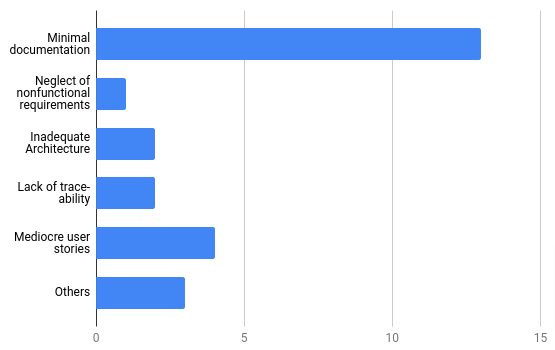}}
\caption{Number of tools/methods resolving a particular challenge}
\label{figure:barcharttoolsolutions}
\end{figure}

Agile methods or tools that have tried to address the challenges in dynamic contexts have gained much interest among practitioners and researchers \cite{1607376}. Keeping that in mind, different researchers attempted to identify those challenges and built tools that provide on-demand solutions \cite{7872736}. We listed all agile challenges in table \ref{table:toolsolutions} that our previously mentioned tools and methods attempted to resolve. Figure \ref{figure:barcharttoolsolutions} represents a number of tools/methods that resolved a particular challenge.

\begin{table*}[htbp]
\caption{The solution provided by tools and methods to certain challenges}
\label{table:toolsolutions}
\begin{center}
\def\arraystretch{1.3}
\begin{tabularx}{\textwidth}{S R B B B}
\hline
\textbf{No.} & \textbf{Challenge} & \textbf{Description} & \textbf{Tools/Methods} & \textbf{Solutions} \\
\hline
1 & Minimal documentation \cite{4420071, 6613063, 7320450} & The lack of documentation imposes a variety of problems including the inability to scale the software and add new members. \cite{4420071} & Source code based documentation (simul loco \cite{6263161}, doxygen \cite{4151761, 10.1007/978-3-642-13881-2_7}, JavadocMiner \cite{10.1007/978-3-642-13881-2_7, 10.1007/978-3-642-21043-3_49, KHAMIS201319}, GitHub plus markdown \cite{8804485}, Graphical UML class models \cite{7967949, 6298089, 6832301, 7059146, 8449554, 5513837, 6895218, 8782418}, XML\_DocTracker \cite{7885872}), Wiki based (Wikis \cite{ambler2002agile, silveira2005wiki, 5978877}, sprintDoc \cite{7521550}, XSDoc \cite{7521550, 10.1145/1104973.1104980, aguiar2003xsdoc}), Flow chart or Diagrams (Flowgen \cite{6975637}, CLARET \cite{8491148}, Ticket-commit network chart (TCC) \cite{8048905, saito2018discovering}), Abstract specification tool \cite{6614746} & Generates minimal documentation with a very little effort. The documentation can be either generated from the source code comments or XML or YAML or JSON file. \\
\hline
2 & Neglect of non-functional requirements \cite{4420071, 7113503, 7133007, 7133028, 7320450} &  Customers often prioritize core functionality over non-functional requirements (NFRs) such as scalability, maintainability, portability, safety, or performance. \cite{4420071, 7925365} & TraceMan \cite{6824127} & Enables traceability among non-functional requirements. \\
\hline
3 & Inadequate Architecture \cite{ramesh2010agile, 6324602} &  As further requirements are known, the architecture designed by the development team during the initial phases may become outdated or inadequate. \cite{ramesh2010agile} & Architectural (Abstract specification tool \cite{6614746}, Active Documentation Software Design (ADSD) \cite{Rubin2011}) & Generate architecture design with relevant and updated information. \\
\hline
4 & Lack of traceability \cite{Cleland-Huang2012, 9263420, 6716450} &  Tracing becomes challenging when dealing with large scale or distributed software development efforts. \cite{Cleland-Huang2012} & Traceability (TraceMan \cite{6824127}, Trace++ \cite{7765512}) & Trace appropriate user stories or different artifacts even if the product scales in the future. \\
\hline
5 & Mediocre user stories \cite{8491182, 8560942, 7113503} & Lack of proper user stories leads to failure to achieve an overview of the product for team members. \cite{8491182, 8864199} & User stories based (COSMIC method \cite{6113071, 7000106}, CMMI \cite{6968220, 6209192}, LAQF \cite{8966716}), TraceMan \cite{6824127} & Measure the quality of user stories and help to write high-quality user stories. \\
\hline
6 & Others \cite{Cleland-Huang2012} &  & API/Web Service (Docio \cite{7961538}), View based (View-based software documentation \cite{1607376}), Scrum (Scrumconix \cite{7577440}) & Generate API documentation from JSON file or supports view based software documentation. \\
\hline
\end{tabularx}
\end{center}
\end{table*} 

\subsubsection{Minimal documentation} 
One of the primary challenges in agile documentation is to keep the documentation minimal \cite{4420071, 6613063}. 
Many documentation generation tools that generate documentation from source code and chart, diagram, and flowchart-based documentation evolved to keep documentation minimum and simple. Practitioners must keep minimal documentation to enhance agile software products, and the tool simul loco comments may answer this problem. Simul loco documentation is extremely useful \cite{6263161}. Moreover, GitHub plus Markdown support options so that reviewers can give a quick review, and it only takes a few minutes for a minor update. A document can be improved easily and continuously \cite{8804485}. Abstract specification tool proposes a considerably shorter abstract specification document, requiring minimal documentation efforts and resulting in shorter documentation that is easier to review, update, and communicate \cite{6614746}. 

\subsubsection{Neglect of non-functional requirements} 
The effect of requirements changes on the architecture is crucial. It was difficult to trace precisely which architectural decisions had to be reconsidered because of the lack of traceability between the textual requirements specification documents and the architectural models. TraceMan fixes this since the trace links are created during the artifacts’ creation. As a result, we can better understand the functional and non-functional requirements using TraceMan more accurately and consequently \cite{6824127}.

\subsubsection{Inadequate Architecture} 
The primary goals of agile development are flexibility, minimalism, and collaboration. Abstract specification tool achieves these by creating a short and focused architecture document \cite{6614746}.

\subsubsection{Lack of traceability}
Conventional agile projects entail intensive labor work to generate and maintain traceable links. Consequently, lack of traceability provides a weak layer over the software system no matter how much flexible the system is \cite{Cleland-Huang2012}. On the other hand, trace++ generates a large number of traceability relations combining the various artifacts \cite{7765512}.

\subsubsection{Mediocre user stories} 
Although user stories are essential for ASD, people struggle to document high-quality user stories. Even user stories mentioned in the current dataset are of poor quality  \cite{8491182}. TraceMan produces high-quality user stories by having detailed traces of user store information \cite{6824127}. 

\subsubsection{Others} 
Some tools resolve the complex API documentation challenges by creating API documentation with ease \cite{7961538}. In addition, there are some tools such as Scrumconix \cite{7577440}, and view-based software documentation that cover challenges posed in the area of scrum and view-based \cite{1607376}.


\section{Threats to Validity}
\label{sec:Threats}
The threats to our systematic literature review are the specification of the candidate pool of papers and primary study selection bias.  We selected our primary pools of studies through database searches and used keywords. Our keywords were very precise, and we obtained a good number of papers. However, we may missed some papers due to our specific search string. 

We also used a specific period to select our studies, which might have discarded relevant papers. On the other hand, we relied on IEEE Xplore for a primary pool of studies, which threatens to have a complete set of primary studies. To mitigate this risk, we performed both backward and forward snowballing, which eventually resulted in a collection of papers from other databases like ACM, Springer, etc. We followed the standard inclusion and exclusion criteria, which might still introduce some personal bias. 

\section{Conclusion}
\label{sec:Conclusion}
Working software gets priority over detailed documentation in agile software development. Even though documentation is less of a priority in ASD, studies have shown that a minimal level of documentation is essential. This research aimed to identify key elements to record in ASD and locate appropriate tools to aid in documentation. We conducted a systematic literature review to identify essential information in agile documentation and the effectiveness of current methodologies and tools in agile software development. As a result, we have compiled a list of essential elements in agile documentation and tools and approaches that can help alleviate the documentation challenges.

Our findings will aid in understanding key aspects of agile documentation and how agile documentation tools and approaches function. In the future, we want to map the relationships between these technologies and develop a method that can be used as a one-stop solution for agile documentation. We also intend to conduct a multi-vocal literature review to find more industry concerns and solutions. Finally, our future plan also involves a survey of agile practitioners to see the usefulness of this article.


\bibliographystyle{abbrv}
\bibliography{sigproc}

\end{document}